\documentclass{article}

\usepackage{arxiv}

\usepackage[utf8]{inputenc} 
\usepackage[T1]{fontenc}    
\usepackage{hyperref}       
\usepackage{url}            
\usepackage{booktabs}       
\usepackage{amsfonts}       
\usepackage{nicefrac}       
\usepackage{microtype}      
\usepackage{lipsum}
\usepackage{color} 
\definecolor{mygreen}{RGB}{28,172,0} 
\definecolor{mylilas}{RGB}{170,55,241}

\usepackage{amsthm,amssymb,amsmath}
\usepackage{longtable}
\usepackage{geometry}

\newtheorem{lemma}{Lemma}

\usepackage{algorithm}
\usepackage{algpseudocode}
\usepackage{listings}

\usepackage{tocloft}
 
\title{Assortment Optimization with Customer Choice Modeling in a Crowdfunding Setting}
\author{ Fatemeh Nosrat, 
\\\\
\textit{Department of Industrial and Systems Engineering, University of Minnesota}
\\\\
\textit {
nosra002@umn.edu,
}
}

\begin{document}

\maketitle

\begin{abstract}
Crowdfunding, which is the act of raising funds from a large number of people's contributions, is among the most popular research topics in the economic theory. Due to the fact that crowdfunding platforms (CFPs) have facilitated the process of raising funds by offering several features, we should take their existence and survival in the market place into account. In this study, we investigated the significant role of platform features in a customer behavioral choice model. In particular, we proposed a multinomial logit model to describe the customers' (backers') behavior in a crowdfunding setting. We proceed by discussing the revenue sharing model in these platforms. For this purpose, we conclude that an assortment optimization problem could be of major importance in order to maximize the platforms' revenue. We were able to derive a reasonable amount of data in some cases and implement two well-known machine learning methods such as the multivariate regression and the classification problems to predict the best assortments the platform could offer to every arriving customer. We compared the results of these two methods and investigated how well they perform in all cases. 
\\
\\
Hereinafter, for avoiding unnecessary repetition of words, we use project (owner), campaign (owner), and product (owner), alternately, in the paper. In addition, backers and customers are considered the same individuals.
\\
\\
\textbf{Keywords}: Crowdfunding, crowdfunding platforms, investment- and reward-donation-based crowdfunding, assortment planning optimization.
\\
\\
\textbf{Category}: q-fin
\end{abstract}

\section{Introduction}
One of the outcomes of the 2008 global financial crisis was the “funding gap” which could be defined as a situation where banks were reluctant to provide loans and investors moved capital into more stable platforms. By and large, raising capital became the greatest challenge for small and midsize businesses in this circumstance due to the fact that the market could not tolerate high risk which then leads to the emergence of new crowdfunding platforms. This phenomenon could further be supported by \textit{endogenous growth theory}. This economic theory asserts that investments in human capital, innovation, and knowledge are significant contributors to economic growth.
\\
\\
Crowdfunding developmental efforts entered a new stage in the U.S. when President Obama signed the JOBS Act on April 5, 2012. Notwithstanding the Securities and Exchange Act of 1934, when only accredited investors had the ability to invest in private equities, the JOBS Act opened the door for entrepreneurs to access an entirely new source of capital from non-accredited investors. The new law allows non-accredited investors to invest up to $2,000$ per company every $12$ months, giving them the opportunity to participate in economic growth while supporting the companies they want to support. Consequently, crowdfunding became a new legal source of capital for newly arrived businesses. While experiencing exponential growth in popularity in 2012, the number of crowdfunding websites in the U.S. has raised up to more than 50 at the time (Egerber, Juliehui, and Pykuo 2012).
\\
\\
Taking the Job Act into account as a milestone in the context of investment and financing, there are several reasons why building crowdfunding platforms are inevitable afterwards. First, matching an enormous crowd of funders with a large number of project owners is now more secure, efficient, and effective due to low online search costs. Second, enabling funding in small increments is economically feasible via the internet, and thereby, diminishes the risk exposure. Third, low communication costs facilitate better information gathering and progress monitoring for distant funders.
\\
\\
Regarding the paramount role of CFPs, we will delve into the key aspects that could lead to the success of crowdfunding efforts. First and foremost, CFPs are meaningless without their participant individuals. Therefore, a motivational analysis from both creators' and funders' perspectives seems crucial. Second, personal networks and the underlying projects' quality features are significant factors in a thriving platform. Finally, a platform's business model mainly impacts its progress and developmental procedure. In fact, a business model encompasses two main models denoted as supply and revenue models. In this case, we need to pay heed to the revenue model of these platforms, and thereby, our goal would be to take revenue maximization as an objective to further investigate on.
\\
\\
The revenue maximization problem could be discussed from different perspectives. We may assume that the platforms revenue consist of the following parts:
\\
\\
1. Transaction fee: when people back (donate to or buy) projects 
\\
2. A percentile of funding goal: If the project achieves its goal (succeed) 
\\
3. Service fee: If projects are being put in the first pages or in a way to get more attention, the platform will charge a higher fee. On the other hand, projects with higher popularity and contribution rate will have lower service fee ( since putting them in first pages would be beneficial for the platform as well)
\\
4. Timeline fee: If the project owner wants to extend the deadline of raising fund, the platform will charge an extra fee. 
\\
\\
Knowing the above partition, it is easy to verify that section three is the only segment of the revenue model which could be governed by the platform. Now, the question would be to recognize those platform features that could be important in this part. In this case, we assumed that there exist an assortment optimization problem for the platform to be solved in order to attain the highest possible benefit. Thus, we tried to define the influential parameters within the model, and then solve the relative optimization problem in this study. 
\\
The aforementioned assortment optimization problem differs from other assortment optimization problems due to its specific features. For this purpose, we compared assortment problems in some regular platforms such as Amazon with our problem instance such as IndieGoGo. The comparison further revealed the importance of devoting time and effort to this topic.  

\section{Literature Reviews}
\label{sec:headings}
The financing of early-stage creative projects and ventures was typically based on personal relationships and required face-to-face interactions. Moreover, this traditional procedure involves high levels of risk, uncertainty, and information asymmetry (Agrawal et al. 2014). The aforementioned issues lead to the recent rise of an innovative methodology to raise funds which are being regarded as "Crowdfunding". In fact, one of the most striking features of "Crowdfunding" is the broad geographic dispersion of both creators and investors (Agrawal et al. 2011). Furthermore, there are some hidden information problems since funders may lack the necessary information to estimate the chances of success of a project. Also, some hidden action problems may arise due to funders not being able to control how fundraisers utilize the provided funds (Belleflamme, Omrani, and Peitz 2015).
\\
\\
Our goal in this paper was mainly to understand the concept of crowdfunding and the idea behind the current crowdfunding platforms, and thereby, propose a model that can capture most of the influential factors within these platforms. In addition, we proceed by highlighting the possible extensions of our works as well as some paramount existing research problems and questions in this field.
\\
\\
To define the concept of crowdfunding, we began by investigating the meaning of this term which could typically be found in well-known dictionaries. Crowdfunding could be defined as the practice of gaining needed funding for a project or venture by raising many small amounts of money from a large number of people via the internet (merriam-webster.com, oxforddictionaries.com, retrieved May 2019). However, the idea behind this concept might need some deeper consideration. In other words, crowdfunding is a collective effort by people who network and pool their money together, and therefore, may be affected by or influencing other people or parties. Furthermore, crowdfunding enables easy accessibility of vast networks of people through online platforms by bringing individual investors and entrepreneurs together.
\\
\\
Crowdfunding can be seen as an open call to provide financial resources which mostly takes place on crowdfunding platforms (CFPs). A CFP is a class of two-sided platforms, which provides a matching service between two sides of a market which are defined as fundraisers and funders. Since projects require more than one funder, CFPs provide one-to-many matching systems. In addition, within a CFP, projects have different features and funders have heterogeneous preferences over these projects. Moreover, regarding personal monetary benefits of the funders, crowdfunding platforms could be divided into two main categories denoted as Reward-based CFPs and investment-based CFPs. While Reward and donation-based CFPs fall into the first category, equity-based, Royalty-based and lending-based CFPs could all be considered as the later one (Belleflamme, Omrani, and Peitz, 2015).  
\\
 \\
Exploring the existing literature, we have been able to categorize them into three segments. The first segment would be papers that have empirically investigated the herding behavior caused by asymmetric quality information and observational learning from early contributions (e.g., Agrawal et al. 2011, Freedman and Jin 2011, Zhang and Liu 2012, Mollick 2014). The second segment mainly consists of papers that are based on economic theory and trying to define or modify a framework for different types of crowdfunding. These papers usually involve a detailed investigation on current crowdfunding business models and try to elaborate on contributing factors in the rise of different CFPs (Belleflamme, Omrani, and Peitz, 2015, Agrawal, Catalini, and Goldfarb 2014). The third segment would be some modeling papers consisting of some major assumptions to simplify the system in order to derive some decision policies or optimal designs for these systems. As we can see in (Hu, Li, and Shi 2015), by isolating the effect of asymmetric information, we can study specific marketing decisions under the provision point mechanism, as compared to the conventional selling mechanism. The fourth category, which is not prevalent, involves the concept of \textit{motivational crowdwork} and will utilize data and analysis to reveal the incentives for each party (Egerber, Juliehui, and Pykuo 2012). 
\\
\\
Although we can find the aforementioned four categories of papers within this area, we still lack papers that can combine the later three types. In other words, developing a mathematical model utilizing optimization skills while considering the economic theory, platform features, motivations, and incentives, etc. at the same time, could be an innovative area of research.
\\
\\
This paper has connections to the growing body of literature on assortment optimization models in revenue management. (Talluri and van Ryzin 2004) study assortment optimization problems under the assumption that the parameters of the multinomial logit model are deterministic and known. Under this assumption, the authors show that the optimal assortment includes a certain number of products with the highest revenues. Throughout the paper, we refer to such assortments as revenue-ordered assortments. In this case, the optimal assortment can be obtained by checking the expected revenue from each assortment that includes a certain number of products with the highest revenues. (Rusmevichientong et al. 2013) investigate the question of what we can say about the performance of such assortments when the choice model parameters are random. (Bront et al. 2009) show that an assortment optimization problem with random parameters is NP-complete when the number of possible realizations of the choice model parameters is at least as large as the number of products. In this paper, we consider a different assortment optimization problem to include pivotal factors are unique in a crowdfunding setting. Above papers in the literature study the assortment optimization without having any parameters showing the sensitivity of projects network effects. However, in online crowdfunding settings, all campaigns exhibit network effects; backers know about the number of previous supports, and this information can affect the their utilities of projects. Hence, the network effects of projects are one critical factor which the platform needs to consider when he/she is making decisions about assortments of the projects.

\section{Crowdfunding models and Assortment Optimization}
\subsection{Funding Models in a crowdfunding setting}
One of the funding models which have been utilized in many platforms is threshold-pledge (aka all-or-nothing funding model), in which each campaign is given a fixed funding goal within a stated period of time. If the raised fund at the end of the given period is less than the goal, the campaign fails, and no money will flow from funders' credit cards to fundraisers - and the platform earns nothing. Evidence shows that more than 50 percent of CFPs have implemented such a system. Taking platforms with this system into account, when the funding target is reached, the platform allows funders to join some but not all campaigns for other additional funding until the deadline is reached. 
\\
\\
Another funding model implemented by crowdfunding platforms is "keep-it-all". In contrast with the all-or-nothing model, this method allows campaign owners to keep the funds they raise even if the funding goal is not reached.

\subsection{Types of Crowdfunding models}
Due to the personal monetary benefits of the funders, crowdfunding platforms are being divided mainly into two types which are investment-based CFPs and Reward-donation-based CFPs. 

\subsubsection {Investment-based crowdfunding platforms}
In this kind of platforms, funders make investments to a campaign in order to obtain monetary benefits from that. 
\begin{itemize}
    \item \textbf{Equity-based crowdfunding platforms:} These platforms are typically being used to raise money to fund the launch or growth of a company (not just initiate a creative project). These companies often raise money from angel investors or venture capitalists. In return, investors gain benefits through getting a small piece of equity in the company as a result of their contribution (which is at least \$ 1,000  and often a lot more).
    
\item \textbf{Loyalty-based crowdfunding platforms:} Unlike equity-based CFPs, royalty-based CFPs offer backers a percentage of revenue from the project or venture the backer supports, once it generates adequate capital. 
In other words, the fans who bet on a specific product or company get royalty rewards in the future and this could be considered as a type of revenue-sharing.
Moreover, there is a variation for loyalty-based approach in which fans support a project by pre-ordering the product. 

\item \textbf{Lending-based crowdfunding platforms:} 

In these platforms, Lending-based CFPs funders are offered a certain interest rate on successful projects if the project succeed(i.e achieved its funding goal). Indeed, lending can bypasses traditional banks due to the previous fact. In addition, unlike traditional banks, the CFP does not screen between different projects. Instead, it will allow the funders to decide for themselves if a particular project should be funded or not.
An interesting feature of Prosper is that it allows funders to organize themselves in groups. For this purpose, each funder can set up a group and act as a group leader. As the leader, this funder can recommend and comment on particular campaigns. Also, investments can be made public within the group. The group leader is allowed to charge for his or her services. Lending Club is the world’s largest peer-to-peer lending platform. It enables borrowers to obtain a loan, and investors to purchase notes backed by payments made on loans. Based in San Francisco, California, USA. Zopa is the world’s oldest and Europe’s largest peer-to-peer lending service having now lent over £1 billion. Zopa peer-to-peer lending works by bringing together individuals who have money to lend, and individuals who wish to borrow money. 

\end{itemize}

\subsubsection {Reward-donation-based crowdfunding platforms}

In the second class, funders cannot expect monetary compensation. They fund a campaign because they obtain a product or because they support its cause (or a combination of the two). Crowdfunding platforms in this category do not offer a stake in the project or a monetary payment. They offer rewards to funders instead. For instance, a campaign attempting to fund a new video game may consider rewarding each funder a free copy or a personalized version of the game. Even donations in this type of crowdfunding platforms can be personally beneficial when platforms mention the funders' names as donors. 
\\
\\
The first reward-donation-based crowdfunding platform is ArtistShare. It performed successfully by raising an acceptable amount of fund from a large number of people for artists. ArtistShare's success provided a strong incentive for launching more reward-based crowdfunding platforms such as IndieGoGo in 2008 and Kickstarter in 2009 to host funding campaigns with a wider variety of projects. For example, individuals with such social causes as animals, community, education, environment, health, politics, and religion as well as entrepreneurs and small businesses with projects on food, sports, gaming, publishing, and technology now are given a golden opportunity to raise funds on these platforms.

 \begin{itemize}
     \item \textbf{Kickstarter} This reward-donation-based crowdfunding platform implemented an all-or-nothing funding model, where the campaign's success relies on whether the raised fund at the end of the period is above the funding goal or not. It is obvious that all projects cannot be funded, of course. 36 percent of Kickstarter projects are reported to be fully funded based on their stated goals and deadlines, while the majority walk away with nothing. 
     \item \textbf{IndieGoGo}. This is another reward-donation-based crowdfunding platform, where campaigns are mostly about innovative ideas and technology. In contrast with Kickstarter, campaigns on these platforms can use either the all-or-nothing or keep-it-all methods to fund their projects. Successful campaigns are also able to sell their items in the market through the designed marketplace of the platform. We discuss this platform in more details in the next section.
 \end{itemize}
 
Considering all of the above crowdfunding models, 

\section{IndieGoGo}

IndieGoGo is an international online crowdfunding platform, where project owners and start-up firms can introduce their innovative interesting ideas to all people around the world and raise funds in order to turn their great ideas into a product and create their own markets. Then, the project owners, who have their own markets and a good source of customers, can enter to the competitive atmosphere of marketing and sell their products to a broad group of customers in all over the world. IndieGoGo is one of the first online crowdfunding platforms which allows people with different races, genders, level of education, to name but a few, to have an opportunity to share their innovations with all groups of people (investors, diverse income-level people, etc.) and ask them for funds directly.
\\
\\
\textbf{Differences with other online platforms} First, there is a high chance of fraud on IndieGoGo. Campaign owners can reach their goals, break their commitments, and leave the platform and their backers without producing and shipping any of their products. The chances of these kinds of frauds are almost zero on Amazon. second, campaign owners cannot prepare their ideas in a short amount of time and the delivery dates are not accurate. They provide the backers with estimation dates, which are in month/year basis, and it is not clear exactly when the products are ready and will be received. In addition, those estimation dates can be totally changed, because there is a high chance that the campaign owner cannot reach his/her goal (raise the needed fund), the customers waste their time waiting a long time for the product. Amazon, on the other hand, provides each customer with the roughly exact delivery time, which is short and convenient. Since Amazon and its third parties are able to produce and prepare their products without the need for funds, the waiting times for the customers are extremely lower than the ones in IndieGoGo. Last but not least, the ideas can be stolen by companies located in other countries and be produced at a cheaper price, and therefore, the campaign owners lose their ideas. The occurrence of events such this is rare on Amazon.
\\
\\
\textbf{Fees and Revenues.} Any campaign owners are not charged by the IndieGoGo platform when they start using the platform to advertise their ideas. When the campaign owners reach their goals and raise the fund, IndieGoGo then will ask the campaign owner to share $ 5\% $ of their contributions. This method of charging the campaign owners is known as the revenue sharing in management science and economics. The campaign owners will also be charged $ 3\% + \$0.30 $ per transaction fees for credit card processing. 
\\
\\
Our motivation in this topic could correspond to the unique features associated with the crowdfunding platforms such as their revenue models that differ them from conventional ones presented in the market. For one thing, some levels of uncertainty exist in terms of the products' availability, leading time, to name but a few. Moreover, these platforms are free of charge for public users and gain their profits from a revenue sharing context defined between platforms and project' owners. In this case, the profits would only be generated if the owners attain their predefined funding goals, and therefore, both parties equally put efforts to succeed. 
\\
\\
Although campaign owners are striving to raise funds for their projects, because of the special characteristics of the revenue model, in this case, the platform's suggested assortment could play a pivotal role in the success of both parties. 
\\
\\
Generally, product assortment selection is among the most critical decisions in each platform. Thus, one of the central decisions would be which products to include in the platform's project assortment. This will be referred to as an assortment planning problem.

Next, I start discussing the methodologies needed to use to formulate the problem.

\section{Model Description}  

We suppose that the platform provides a more user-friendly environment for backers by customizing itself with a search engine to help backers find what projects they wish to support. The platform also implements several filter options for refining the search results because these exist a wide variety of innovative projects in the system and are strongly influenced by the inter-industry technologies. Hence, eventually, backers are assigned to a more focused group of projects in the same or other industries and cannot be distracted by fancy projects. Therefore, no unexpected purchase behavior can be observed in the system. 
\\
\\
However, it is not realistic to restrict the backer to that degree of search filtering in the system. For omitting that constraint on searching processes, by convention, we assume backers considers utilities very close to zero for those unrelated projects.
\\
\\
Even though the customers will reach pages with more related campaigns, since the number of these campaigns, which needed to be funded in the platform, is still absolutely tremendous, they probably face with so many pages of results contained different campaigns. However, the customers logically do not consider all pages so as to find the best-fit campaign(s). Almost all of the time, customers have a limited period of time available to look over among the campaigns may skip all of them except those on a few first pages, and if they do not find the campaign will leave the website without even check the other pages of the results. By taking this backer's behavior into account, the platform should decide what assortment of projects offer on these pages in order to maximize its revenue. 
\\
\\
We assume that the number of the campaigns which can be shown on the first page of the search results is at most $ n $. Every time customers search the word, the platform owner should decide what assortment of the products (a subset of $ n $ products) offers to the customers in order to maximize its revenue from each customer. Note that this assortment is different for each customer because customers' valuations of the projects are different and optimal assortment for one customers may not be optimal for another customer who has a different valuation of the projects. 
\\
\\
The factors can have the influence on the valuations of the customer for each customer are the information about the levels of funds, which have been raised so far, the number of available perks to purchase, and remaining time for reaching the funding goal. We consider some of these factors in our formulations.
\\
\\
By considering all the above discussions, we describe our problem with following assumptions.
\\
\\
We consider a crowdfunding platform faced by a problem which involves choosing an assortment of available projects with networks effects to offer to their backers with the goal of maximizing revenues from each of them.
\\
\\
In addition to above assumptions on the customers' utilities of the campaigns, there are also some other assumptions needed to be made on the problem, which describe the special characteristics of the projects, the types of customers, and the factors which cause changes in the customer's utility of each project for better determination of the platform processes in this setting. The reasons behind making all these assumptions are, first, the projects may serve as substitutes and the customers may choose among the projects that satisfy their needs or wishes. This creates a situation where the demand for each project depends on what assortments of projects are offered to the customers. Therefore, one has to consider the preferences of the different customer segments, as well as the size of each segment, when deciding which assortment of projects to offer. Second, nowadays, most of the projects have different features which can help to increase the level of their sales even when the prices of those projects are high. One of these features is the networks effect.
\\
\\
Briefly, a product exhibits network effects if the value that an individual backer obtains from it increases in the number of other backers who also back that project. Incorporating network effects in the backer's utilities can make the problem more complex, although it is undeniable to not consider network effects as an effective factor on the backers' utilities of the projects. There are also some other factors such as the intrinsic value of the projects, the goal of each campaign, the available time to raise fund, the kind of funding (flexible or fixed), heterogeneity in the offering perks, which cause heterogeneity in projects and need to be considered in the model. However, the more factors we consider, the more complicated our model becomes.
\\
\\
As we mentioned above, we have two kinds of funding on the platform: Flexible and fixed. We consider one of the extreme situations of only flexible funding allowing here. When the campaigns are allowed to have a flexible funding goal, we can provide a static model to obtain the optimal assortment. The reason for using this model is that the platform receives a revenue after each support without having to wait for the campaign to reach its funding goal. 
\\
\\
Consider the crowdfunding platform.  
A backer searches over all the projects existed in the system so as to find that project in his/her mind and based on that receives pages of available projects. The customer then considers a utility for each of those projects on the first page, and choose one of them with the maximum utility or leave the system without giving the projects any support.
\\
\\
After the backer chooses the project with the maximum utility, hinge on gaining more information provided by the campaign owner about the features of the chosen project, he/she then either becomes a buyer with a probability and buy one of the available perks, or will become a donor with another probability and donate the project, without purchasing any of those perks. By convention, we suppose that choosing a project is equivalent to support that project by either purchasing or donating. Note that this assumption does not restrict the problem a lot because these platforms are not marketplaces, where backers buy anything. The backers can donate any amount of money, and if they back a project above a threshold, then they can receive one of the perks as a reward/gift of their donations.

\subsection{Technical Approach}

A standard approach to modeling demand in revenue management is to assume that each customer arrives into the system with the intention of supporting a fixed campaign. If this campaign is available for support, then the customer support it. Otherwise, the customer leaves the system without making a contribution. In reality, however, there may be multiple projects that can potentially serve the needs of a customer, in which case, customers may make a choice between the campaigns and may substitute a campaign for another one when their favorite campaign is not available. This kind of a choice process creates interactions between the demand for the different projects, increase the demand for an available project when some other projects are not available so that customers satisfy their needs by substituting for the available projects. A question that arises in this setting is what campaigns should be available to customers so as to maximize the expected revenue from each customer, given that customers choose and substitute according to a particular choice model. We consider assortment problems under a \textbf{mixture of multinomial logit models}. There are multiple customer types. Customers of different types choose according to different \textbf{multinomial logit models} whose parameters depend on the type of the customer. The goal is to find a set of products to offer so as to maximize the expected revenue obtained over all customer types.

\subsection{The General Model - Flexible Funding}

Let consider that there are $ n $ projects indexed by $ 1, 2,\ldots, n $ and there are m segments of customers.  In this case, part of the campaigns' and platform's revenue is guaranteed because they will receive their conventional revenues after customers back the campaigns. Consequently, there is no failure in this model of funding, but the more funds are raised, the more revenues both parties get. Also, from backers' perspective, the campaign's success is more confidently obtained, since the campaign has almost the needed money to launch its new items into the market. We suppose that the platform is received the corresponding revenue of $ r_{ij} $ for contribution/donation from customers in segment $ j $ on project $ i $. For each $ i $ and $ j $, $ r_{ij} $ is a uniformly distributed random variable in the range of $ [a,b] $, where $ a $ is small enough and $ b $ is large enough that can contained all conventional percentage of the donations for platform's revenue. Furthermore, we assume that the platform receives $omega_{ij}$ portion of each contribution from customers in segment $j$ backing project $i$ as well as $\xi_{i}$ portion of all raised funds from the whole population of backers.
\\
\\
The customers in each segment choose among the offered products according to random utility maximization. In particular, for segment $ j $, $ j = 1,\ldots,m $, customers associates the utility $ U_{ij} = V_{ij} + \epsilon_{ij} $ with project $ i $, where $ \epsilon_{ij} $ is a standard Gumbel random variable with mean zero, and we view $ V_{ij} $ as the mean utility of product $ i $ in segment $ j $. We normalize the utility of the no purchase option to zero. In this case, if we offer the assortment $ G \subseteq \lbrace 1,...,n \rbrace $ of products to the customers, then a customer chooses the product with the highest utility if his/her utility of that product is positive, but otherwise, leaves the system without purchasing anything. It is a standard result in a discrete choice theory that if we offer assortment $ G $ to customers in segment $ j $, then a customer in this segment chooses product $ i \in G $ with probability
\[
P_{ij}(G,V) = \frac{\exp(V_{ij})}{1+\exp(V_{ij})},~ i=1,\ldots,n,~ j = 1,\ldots,m,
\]
where we use $ V  $ to denote the vector of mean utilities of the products, and make the dependence of $ P_{ij}(G,V) $ on $ G $ and $ V $ explicit. The choice model above is known as the multinomial logit model. If we offer the assortment $ G $ and the vector of mean utilities is $ V $, then the expected revenue for one customer is
$ h(G,V) $, which is formulated later in this section. When the mean utility vector $ V $ is fixed and known, we can find an assortment that maximizes the expected revenue from a customer by solving the problem $ \max_{G \subseteq \lbrace 1,...,n \rbrace } h(G,V) $. The assumption behind using a fixed mean utility vector is that each customer is reacting to an offered assortment in the same manner. On the other hand, if each customer has a different reaction towards an assortment, then the mean utilities that a customer attaches to the products can be modeled as random variables themselves. In that case, we can solve
\[
W^{*} = \max_{G \subseteq \lbrace 1,...,n \rbrace }\mathbf{E}\lbrace h(G,V)\rbrace,
\]
to find an assortment that maximizes the expected revenue over all possible realizations of the mean utility vector. The above expectation involves the random vector $ V $ and the distribution of $ V $ naturally depends on the composition of the market and how customers in the market make a choice. The random vector $ V $ can have a discrete or continuous distribution, and we assume that it is independent of $ \epsilon $ vector. For simplicity, we suppose that  $ V $ is a discrete random variable with two different values. In other words, we have $ m $ different segments of customers (We can divide customers into different segments by several external factors such as their geographical regions. $ V $ can be formulated as below:
\[
V_{ij} = y_{ij} -\beta_{ij} F_{i} +\alpha_{ij}\sum_{j=1}^{m} \lambda_{j}  q_{ij},~ i=1,\ldots,n,~ j = 1,\ldots,m,
\]
where $ y_{ij} $ is deterministic and represents the intrinsic utilities of products $ i $ for customers in segment $ j $, $ F_{i} $ gives the the goal funding of the product $ i $, $ q_{ij}$ is a decision variable which represents the fraction of customers in segment  $ i $ who purchase product $ i $, $ \lambda_{j} $ is the size of population of segment $ j $, $ \alpha_{ij} $ is the network effects parameter for customers in segment $ j $ buying product $ i $, and $ \beta $ shows how customers are sensitive to difference value of funding goal and the level of contributions for each product so far.  Then,
\[
 P_{ij}(G,V) = \frac{\exp\left(y_{ij} -\beta_{ij}F_{i} +\alpha_{ij}\sum_{j=1}^{m} \lambda_{j} q_{ij}\right)}{1+\exp\left(y_{ij} -\beta_{ij} F_{i}  +\alpha_{ij}\sum_{j=1}^{m} \lambda_{j}q_{ij}\right)},~ i=1,\ldots,n,~ j = 1,\ldots,m,
\]
and the the expected revenue for the platform from each customer is 
\[
 \textbf{E}\lbrace h(G,V)\rbrace = \sum_{i\in G}\sum_{j = 1}^{m} \omega\left(\left(\frac{a+b}{2}\right)  \lambda_{j}P_{ij}(G,V)\right)+\sum_{i\in G}\xi\sum_{j = 1}^{m}\left(\left(\frac{a+b}{2}\right)  \lambda_{j}P_{ij}(G,V)\right),
 \]
 and consequently, the optimal revenue is
 \[
 W^{*} = \max_{G}\sum_{i\in G}\sum_{j = 1}^{m} \omega\left(\left(\frac{a+b}{2}\right)  \lambda_{j}P_{ij}(G,V)\right)+\sum_{i\in G}\xi\sum_{j = 1}^{m}\left(\left(\frac{a+b}{2}\right)  \lambda_{j}P_{ij}(G,V)\right).
\]
We consider a fluid model with a population of size 1, and for simplicity, by convention, we consider $\beta_{ij}=1$. Then, we have  $ q_{ij} = P_{ij}(G,V) $ and
\[
 q_{ij} = \frac{\exp\left(y_{ij} -F_{i} +\alpha_{ij}\sum_{j=1}^{m} \lambda_{j} q_{ij}\right)}{1+\exp\left(y_{ij} - F_{i}  +\alpha_{ij}\sum_{j=1}^{m} \lambda_{j}  q_{ij}\right)},~ i=1,\ldots,n,~ j = 1,\ldots,m.
\]
If $ G^{*} $ be the optimal assortment obtained from the above optimization problem, the platform maximizes its revenue from each customer if it makes available those projects in $ G $ to that customer.
\\
\\
In next section, we check the conditions we need to obtain the optimal assortment in our model. We also investigate the impact of each parameter in the optimal assortment and the corresponding revenue. First, we consider that the network effect sensitivity parameters for all segment are zero. We extend our investigation to the level, at which network effects parameters can affect customers' valuations in each segment for each project.

\section{Numerical Experiment}

In this section, we describe how our data has been generated, how we use the data to use those machine learning methods we have learned in the class and the results of this numerical study. Since the data-driven from IndieGoGo website or any other crowdfunding platforms is not free and we were not able to find a good source of data on the Internet, we considered a variety of randomly-generated problem instances with $m =1,2$ ($m$ shows the number of segments of customers) and $M = 50$ (all parameters except $F_{i}$'s are uniform between $0$ and $M$). For each value of $m$ and $n$ we repeated this process 500 times to generate 500 examples Note that $y$'s and  $\alpha$'s in all these examples are obtained randomly between $0$ and $M$. For producing suitable $\lambda$'s, first, we generate a random column vector with $n$ components, for example, $a$, in which all components are uniform between $0$ and $M$, then $\lambda = a / sum(a)$, which gives us a column vector with all components between $0$ and $1$ and the summation of all components is 1. Also, we considered that the amount of funds each campaign needs to start its venture is uniform in $\lbrace 1\$$,\ldots, $10000\$\rbrace$. For now $\beta_{ij} = 1$ for all $i,j$. For obtaining the data we need the following lemma. By convection, we suppose that the limit of a matrix is the limit of each component of the matrix.
\begin{lemma}\label{lemma1}
Consider the sequence of vectors $ \lbrace {q}^{t}: t = 0,1,2,\dots \rbrace $ that satisfies
\[
q_{ij}^{t} = \frac{\exp\left(y_{ij} -\beta_{ij}F_{i} +\alpha_{ij}\sum_{j=1}^{m} \lambda_{j} q^{t-1}_{ij}\right)}{1+\exp\left(y_{ij} -\beta_{ij} F_{i}  +\alpha_{ij}\sum_{j=1}^{m} \lambda_{j}q^{t-1}_{ij}\right)},~ i=1,\ldots,n,~ j = 1,\ldots,m,
\]
and let $\bar{q}$ be the largest fixed point. 
\begin{enumerate}
    \item[(i)] If $q^0 = \left[ 0\right]_{n\times m}$, then $q^1 \le \lim_{t\to\infty} q^t \le \bar{q}(p)$.
    \item[(ii)] If $q^0 = \left[1\right]_{n\times m}$, then $\ = \lim_{t\to\infty} q^t  \le q^1$.
\end{enumerate}
\end{lemma}
\textbf{Proof of Lemma 1.} Let define
\[
h(q) =  \frac{\exp\left(y_{ij} -\beta_{ij}F_{i} +\alpha_{ij}\sum_{j=1}^{m} \lambda_{j} q^{t-1}_{ij}\right)}{1+\exp\left(y_{ij} -\beta_{ij} F_{i}  +\alpha_{ij}\sum_{j=1}^{m} \lambda_{j}q^{t-1}_{ij}\right)}.
\] 
For part (i), $\frac{\partial h}{\partial q}(q) >= 0$ and $h(q) = q$,and therefore, the sequence $\lbrace q^{t}\rbrace$ in monotone increasing and bounded above by $[1]_{m\times n}$. Hence, $\lim_{t\rightarrow\infty}q^{t}$ exists. The limit is a fixed point, because $h(q)$ is continuous in $q$. Hence, $\lim_{t\rightarrow} q^{t}\leq \bar{q}$, where $\bar{q}$ is the largest fixed point. 

For part(ii), the sequence $\lbrace q^{t}\rbrace$ is monotone decreasing and bounded below by $[0]_{m\times n}$, so the limit in question exists. Also as above, the limit $L:=\lim_{t\rightarrow\infty}q^{t}$ is a fixed point. Thus, $L \leq \bar{q}$. The monotonicity of $h(q)$ in $q$ and $\bar{q}\leq \left[1\right]_{m\times n} = q^{0}$ together imply that $\bar{q}\leq q^{t}$ for all $t$, so $\bar{q}\leq \lim_{t\rightarrow\infty}q^{t} = L$. It follows now that $L = \bar{q}$.\qed

By using \textbf{Lemma} \ref{lemma1}, given any pair of parameters, we are able to compute $q$  and its revenue for any possible assortments, and thus the best assortment could be the one that corresponds to the maximum revenue.
\subsection*{Case 1. Two products and one segment of customer}
Suppose that there exist 2 registered campaigns in the platform, and the platform owner can show 1 of them to each arriving customer. Once a customer visits the website, the platform should quickly decide how to choose the product in order to have its maximum revenue. The revenue model obtained in the previous section for this case is as follows.
\[
q_{i} = \frac{\exp\left(y_{i} -F_{i}+\alpha_{i}q_{i}\right)}{1+\exp\left(y_{i} -F_{i}+\alpha_{i}q_{i}\right)},~ i\in G,
\]
 and the optimal revenue is obtained from
 \[
 W^{*} = \max_{G}\left(\frac{a+b}{2}\right)(\xi+\omega)\sum_{i\in G}  q_{i}.
\]
We use Lemma \ref{lemma1} and our revenue model in this case to obtain some random data. Knowing Lemma \ref{lemma1} can help us to solve the main problem numerically and obtain the maximum revenue and its corresponding assortment. Therefore, given the values of parameters (features), we know what the best assortment is. Hereunder we want to implement the classification problem to see how good its performance is in this and other cases. 

In order to find the best assortment for each arriving customer, we use the classification problem, whose sample data, consisting of 500 samples and 4 features, is obtained and save on data.m. We choose the first 375 data as the training data and the rest of the data as the test data. In our data, the response vector $Y$ is defined as follows:
\[
Y = \left\lbrace\begin{array}{cc}
     (1,0) & \text{if a customer chooses product 1}  \\
   (0,1)  &  \text{if a customer chooses product 2}
\end{array}\right.
\]
We applied multivariate linear regression to estimate the relationship between $Y$ and $X$, i.e., we assumed $Y = b + B X$ and estimated $(b,B)$ using the training set and computed the error rate. 

Let $r_{a}$ and $r_{c}$ represent the Actual Revenue of the problem and the revenue obtained from the classification problem,respectively. We show percentage of loss in revenue by $PRL$ and define as follows:
\[
PRL = 100 \times\frac{\left(r_{a} - r_{c}\right)}{r_{a}}
\]
We realized how good our predictions are and how incorporating different parameters in the model can decrease or increase the maximum revenue of the platform.    
\subsection*{Case 1. Part 1. Two products and one segment of customers with network effects}
In this part $\alpha_{i}$'s can get any random numbers between zero and $M$, but all components of $\alpha$'s cannot be zero.For having access to the data for this case, please see "data.m". The computed parameters of multivariate regression are as follow:
\[
\beta^{1} = \left[\begin{array}{cc}
    0.2561, \\
     0.0097 \\
     0.0075\\
0.2561\\
-0.0085
\end{array}\right], ~~
\beta^{2} = \left[\begin{array}{cc}
   0.3241\\
0.0086\\
0.0026\\
0.3242\\
-0.0083
\end{array}\right], ~~ \text{ErrorRate} = 0.0320
\]

The average of $PRL$ in this case is 2.40\%. Since the error rate and the average of the percentage of loss in revenue are both small enough, we can conclude that when there are 2 products and 1 segments of customers, the classification problem gives us the best assortment in most of the examples, and the platform's revenue is very close to the maximum revenue.

In this part, we also have $\max(r_{a}) = 0.44$, $\min(r_{a}) =  7.2905e-35$, and the average of $r_{a}$ equals $0.3159$.
\subsection*{Case1. Part 2. Two products and one segment of customer without considering network effects}
The problem setting in this case is similar to Part 1 except $\alpha$'s are all zeros, and therefore, the model simplifies as below.
\[
q_{i} = \frac{\exp\left(y_{i} -F_{i}\right)}{1+\exp\left(y_{i} -F_{i}\right)},~ i\in \lbrace 1,2\rbrace,
\]
 and the optimal revenue is obtained from
 \[
 W^{*} = \max_{G}\left(\frac{a+b}{2}\right)(\xi+\omega) \sum_{i\in G}  q_{i}.
\]
We implemented the model by using \textbf{Lemma}\ref{lemma1} and generated 500 different examples of the problem. We will compare the problem in this part (without network effects) and Part 1 (with network effects) to see how much impact this factor has on the maximum revenue of the platform. Note that one of the most important differences between this paper with ones in the literature is including these parameters in the model and its effects on the best assortment for each arriving customer and the maximum revenue of the platform. For having access to the data for this case, please see "data\_p2\_s1\_withoutNWE.m".

When we implemented the classification problem in this case, we get the following values as an estimation of multivariate regression and the error. 
\[
\beta^{1} = \left[\begin{array}{cc}
0.4314\\
0.0048\\
0.4314\\
-0.0095
\end{array}\right], ~~
\beta^{2} = \left[\begin{array}{cc}
0.3427\\
0.0089\\
0.3427\\
-0.0084
\end{array}\right], ~~ \text{ErrorRate} = 0.04
\]
The average of $PRL$ in this case is 1.20\%. Since the error rate and the average of the percentage of loss in revenue are both smaller than the case where the model was incorporated with the network effects parameters, we can conclude that the performance of the classification problem gets better. 

The average, max, and min of $r_{a}$ when we incorporate the network effects parameters are 0.3159, 0.4400, and 7.2905e-35. On the other hand, when we do not consider the network effects parameters the average, max, and min of $r_{a}$ are 0.1990, 0.4400, and 2.8292e-38. 

The average percentage of revenue loss when we change the problem from one with network effect to the another without network effects is about $20\%$, which shows the consideration of network effects parameters in the model is important and beneficial, and in order to have the maximum revenue, we must incorporate this parameter into the model. Otherwise, the platform will lose $20\%$ of its maximum revenue. 

\subsection*{Case 2. Three products and one segment of customer}
In this case, we suppose that there exist 3 campaigns in the platform and the challenge for the platform is to choose one of these three campaigns every time a customer visits the platform. The revenue model we obtained in the previous section can be simplified as the following model. Note that we consider that all arriving customers are behaving in the same way (homogeneous customers) in this case, and as a result, we only have one segment of customers. 
\[
q_{i} = \frac{\exp\left(y_{i} -F_{i}+\alpha_{i}q_{i}\right)}{1+\exp\left(y_{i} -F_{i}+\alpha_{i}q_{i}\right)},~ i\in \lbrace 1,2\rbrace,
\]
 and the optimal revenue is obtained from
 \[
 W^{*} = \max_{G}\left(\frac{a+b}{2}\right)(\xi+\omega) \sum_{i\in G}  q_{i}.
\]
Using the above model and \textbf{Lemma }\ref{lemma1}, we obtained 500 instances of the problem. We implement the classification model and check whether the classification problem can give us a good prediction of the best assortment in each example or not.
\subsection*{Case 2. Part 1. Three products and one segment of customer with network effects}
In this part we assume that $\alpha$'s in the model is a nonzero vector. For having access to the data for this case, please see "data2.m".

Following shows the values of multivariate regression parameters in this part:
\[
\beta^{1} = \left[\begin{array}{cc}
0.1987\\
0.0092\\
0.0077\\
0.1987\\
-0.0093
\end{array}\right],~
\beta^{2} = \left[\begin{array}{cc}
0.2763\\
0.0045\\
0.0052\\
0.2763\\
-0.0092\\
\end{array}\right],~
\beta^{3} = \left[\begin{array}{cc}
0.2003\\
0.0062\\
0.0062\\
0.2003\\
-0.0080
\end{array}\right]
\] 
The error rate and the percentage of loss in revenue are 0.12 and $9.69\%$. Also, $\max (r_{a}) = 0.44 $, $\min (r_{a}) = 7.9430e-30$, and the average of $r_{a}$ is $0.3720$.
\subsection*{Case 2. Part 2. Three products and one segment of customer without network effects}
In this part, we suppose that $\alpha$'s are all zeros. Therefore, the dimension of classification problem decreases. The data for this part is saved in $data$. The parameters of the regression problem are:
\[
\beta^{1} = \left[\begin{array}{cc}
0.2419\\
0.0107\\
0.2419\\
-0.0082
\end{array}\right],~
\beta^{2} = \left[\begin{array}{cc}
0.2652\\
0.0093\\
0.2652\\
-0.0085
\end{array}\right],~
\beta^{3} = \left[\begin{array}{cc}
0.2557\\
0.0074\\
0.2557\\
-0.0078
\end{array}\right]
\] 
Result 1. The error rate and the average of percentage of loss are $0.072$ and $5.42\%$ for the classification problem. The reason that we see better performance of the classification problem is the dimension of the main problem in this case. As the dimension of the problem gets lower, the classification problem performs much better and gives us a good predication of the best assortment.

Result 2. Note that $\max(r_{a}) = 0.44$, $\min(r_{a}) = 1.8762e-36 $, and the average of $r_{a}$ is $0.2594$. Although the maximum of $r_{a}$ is the same as Part 1, the average of $r_{a}$ is lower compared to the one in Part 1. This outputs show that the platform should take into account impact, which the network effects has on the maximum revenue. Without incorporating network effects into the model, the platform could lose a large amount of revenue and it is inevitable factor at least in the revenue model.
\subsection*{Case 2. Part 3. Three products and one segment of customer with network effects}
One of the questions that arise in this case is that what we can say about the best assortment and the maximum revenue when we can offer two campaigns to each arriving customer. We devote this part to answer this question and check every needed output. Note that we include the network effects parameters in the model because we have realized that this is a necessary factor of the problem. The only difference between the revenue model in this part and the problem in Part 1 is that $G$ has two members. All data related to this part can be found in "data2-As2.m". The implementation of the problem in this part is the same as the case, where we offer only one of the three campaigns to each customer except we compute the norm $l_{2}$ of each response vector with one of the vectors $(1,1,0)$, $(0,1,1)$, and $(1,0,1)$, which causes to obtain an assortment of two campaigns out of three.

Following are the parameters of the multivariate regression:
\[
\beta^{1} = \left[\begin{array}{cc}
0.4732\\
0.0057\\
0.0003\\
0.4732\\
-0.0071
\end{array}\right],~
\beta^{2} = \left[\begin{array}{cc}
0.4179\\
0.0095\\
0.0007\\
0.4179\\
-0.0083
\end{array}\right],~
\beta^{3} = \left[\begin{array}{cc}
0.4076\\
0.0097\\
0.0033\\
0.4076\\
-0.0111
\end{array}\right]
\] 
the error rate  and the average percentage of loss in revenue in this part are 0.152 and $19.20\%$. In addition, $\max (r_{a}) = 0.88 $, $\min (r_{a}) = 1.0315e-28$, and the average of $r_{a}$ is $0.5511$.

Result 1. Comparing this part with Part 1, we realize that the performance of the classification problem gets worse as we increase the number of campaigns in the offered assortments, and consequently, the classification problem cannot give us a good predication of the best assortments in this part compared to Part 1. 

Result 2. As the values of $r_{a}$ in this part and Part 1 show, showing two campaigns to an arriving customer is beneficial and even in the worst case, the platform will gain more revenue if it provides an assortment of two out of three campaigns every time a customer visits the platform.
\subsection*{Case 3. Five products and one segment of customer with network effects}
In this case, we increase the number of products to 5 and still consider that the customers are all homogeneous, that is, they all behave in the same toward any kinds of assortments and campaigns. This homogeneous behaviour causes the number of segments in our proposed revenue model be equal to 1.

One reason that we consider higher number of products (for example, 5 products) when there is only 1 segment of customers is to understand the performance of the classification problem and how well it can predict the best assortments. The following model is the revenue model in this case.

\[
q_{i} = \frac{\exp\left(y_{i} -F_{i}+\alpha_{i}q_{i}\right)}{1+\exp\left(y_{i} -F_{i}+\alpha_{i}q_{i}\right)},~ i\in G
\]
 and the optimal revenue is obtained from
 \[
 W^{*} = \max_{G}\left(\frac{a+b}{2}\right)(\xi+\omega) \sum_{i\in G}  q_{i}.
\]
\subsection*{Case 3. Part 1. Five products and one segment of customer with network effects}
Assume that $\alpha$'s is a non-zero vector. The parameters of the multivariate regression problem in this case are as follow:
\[
\beta^{1} = \left[\begin{array}{cc}
0.1311\\
0.0065\\
0.0091\\
0.1311\\
-0.0077
\end{array}\right],~
\beta^{2} = \left[\begin{array}{cc}
0.1030\\
0.0082\\
0.0064\\
0.1030\\
-0.0070
\end{array}\right],~
\beta^{3} = \left[\begin{array}{cc}
0.1250\\
0.0064\\
0.0055\\
0.1250\\
-0.0066\\
\end{array}\right],~
\beta^{4} = \left[\begin{array}{cc}
0.1108\\
0.0056\\
0.0033\\
0.1108\\
-0.0055
\end{array}\right],~
\beta^{5} = \left[\begin{array}{cc}
0.1310\\
0.0048\\
0.0048\\
0.1310\\
-0.0062
\end{array}\right]
\] 
For having access to the data, please see "data3.m". In this case error rate = 0.2 and average of Percentage of loss: 27.69\%. As the results of numerical experiments shows, the error rate and the average of loss in revenue get higher as the number of products increases. In other words, the likelihood that the classification problem gives us the optimal assortment becomes lower if the number of products is large. In addition, $\max(r_{a}) = 0.44$, $\min(r_{a}) = 2.6434e-25$, and $mean(r_{a}) = 0.4176$.
\subsection*{Case 3. Part 2. Five products and one segment of customer without network effects}
The data for this part can be found in "data3-p5-s1-as1-withoutNWE.m".
The average percentage of loss in revenue and the error rate are 8.78\% and 0.0960, respectively. Also,
 $\max (r_{a}) = 0.8800 $, $\min (r_{a}) = 3.5079e-27$, and the average of $r_{a}$ is $0.7399$.
\[
\beta^{1} = \left[\begin{array}{cc}
0.2544\\
0.0095\\
0.0077\\
0.2544\\
-0.0092
\end{array}\right],~
\beta^{2} = \left[\begin{array}{cc}
0.2552\\
0.0098\\
0.0068\\
0.2552\\
-0.0105
\end{array}\right],~
\beta^{3} = \left[\begin{array}{cc}
0.2338\\
0.0096\\
0.00548\\
0.2338\\
-0.0097
\end{array}\right],~
\beta^{4} = \left[\begin{array}{cc}
0.2495\\
0.0093\\
0.0071\\
0.2495\\
-0.0102
\end{array}\right],~
\beta^{5} = \left[\begin{array}{cc}
0.2997\\
0.0084\\
0.0037\\
0.2997\\
-0.0102
\end{array}\right]
\]

\subsection*{Case 3. Part 3. Five products and one segment of customer without network effects}
In this part, we consider that the platform offers any 2 out of 5 campaigns to each arriving customers. All the results are computed by using the data in "data3-p5-s1-2as.m". In this part, we have average $PLR$ = 33.51\%, error rate: 0.1760, $\max (r_{a}) = 0.8800 $, $\min (r_{a}) = 3.5079e-27$, and the average of $r_{a}$ is $0.7399$. Also the parameters of the multivariate regression are:
\[
\beta^{1} = \left[\begin{array}{cc}
0.2544\\
0.0095\\
0.0077\\
0.2544\\
-0.0092
\end{array}\right],~
\beta^{2} = \left[\begin{array}{cc}
0.2552\\
0.0098\\
0.0068\\
0.2552\\
-0.0105
\end{array}\right],~
\beta^{3} = \left[\begin{array}{cc}
0.2338\\
0.0096\\
0.00548\\
0.2338\\
-0.0097
\end{array}\right],~
\beta^{4} = \left[\begin{array}{cc}
0.2495\\
0.0093\\
0.0071\\
0.2495\\
-0.0102
\end{array}\right],~
\beta^{5} = \left[\begin{array}{cc}
0.2997\\
0.0084\\
0.0037\\
0.2997\\
-0.0102
\end{array}\right]
\] 
\subsection*{Case 3. Part 4. Five products and one segment of customer with network effects}
In this part, the implementation of the problem has changed so that the platform can offer only assortments of three out of five campaigns every time a customer visits the platform. The data which has been use for this part can be found in "data3-p5-s1-as3.m". The results of implementation gives are all the following values: average $PLR$ = 52.71\%, error rate: 0.24, $\max (r_{a}) = 1.32 $, $\min (r_{a}) = 1.3555e-19$, and the average of $r_{a}$ is $0.9295$.
\[
\beta^{1} = \left[\begin{array}{cc}
0.3778\\
0.0078\\
0.0024\\
0.3778\\
-0.0062
\end{array}\right],~
\beta^{2} = \left[\begin{array}{cc}
0.4082\\
0.0062\\
0.0044\\
0.4082\\
-0.0091
\end{array}\right],~
\beta^{3} = \left[\begin{array}{cc}
0.4039\\
0.0098\\
0.0040\\
0.4039\\
-0.0111
\end{array}\right],~
\beta^{4} = \left[\begin{array}{cc}
0.3463\\
0.0105\\
0.0063\\
0.3463\\
-0.0115
\end{array}\right],~
\beta^{5} = \left[\begin{array}{cc}
0.3500\\
0.0117\\
0.0048\\
0.3500\\
-0.0114
\end{array}\right]
\] 
\subsection*{Case 3. Part 5. Five products and one segment of customer with network effects}
As the final part of Case 3, we consider that all offered assortments by the platform has 4 of 5 campaigns, and we save the corresponding data in "data3-p5-s1-as4.m". The results of the implementation are as folow:
average $PLR$ = 75.20\%, error rate: 0.2480, $\max (r_{a}) = 1.72 $, $\min (r_{a}) = 1.4160e-24$, and the average of $r_{a}$ is $0.9919$.
\[
\beta^{1} = \left[\begin{array}{cc}
0.4769\\
0.0015\\
0.0019\\
0.4769\\
-0.0022
\end{array}\right],~
\beta^{2} = \left[\begin{array}{cc}
0.4594\\
0.0037\\
0.0012\\
0.4594\\
-0.0032
\end{array}\right],~
\beta^{3} = \left[\begin{array}{cc}
0.5329\\
0.0025\\
-0.0009\\
0.5329\\
-0.0051
\end{array}\right],~
\beta^{4} = \left[\begin{array}{cc}
0.4747\\
0.0069\\
0.0018\\
0.4747\\
-0.0081
\end{array}\right],~
\beta^{5} = \left[\begin{array}{cc}
0.4657\\
0.0111\\
0.0014\\
0.4657\\
-0.0124
\end{array}\right]
\] 
\textbf{Results of Case 3.}
First of all, when we compare Part 1 and Part 2, we can conclude the same the classification problem has a better performance in Part 2 because of the decrease in the dimension of the problem. This result is similar to Case 1. However, ignoring the network effects causes a big loss in the revenue of the platform and for preventing that much loss the platform must include this factor in the revenue model. Now that we have realized the importance of the impacts which network effects parameters have on the revenue, we incorporate this factor into the model in other parts of this case. 

Second of all, when investigating the results of numerical experiments in Parts 3-5, we understand that the classification problem has a poor performance in all parts and as the number of campaigns in each offered assortments gets higher, the quality of classification problem becomes lower. Hence, we expect the classification problem does not give us the best assortments in all examples of the main problem.

Finally, by checking the revenue in Parts 3-5, we understand that the average revenue increases as the number of campaigns in the assortments increases. Therefore, we can conclude that the platform can gain a significant amount of revenue by increasing the number of campaigns in offered assortments of the platform.

\subsection*{Case 4. Two products and Two segments of customers with network effects}
In this case, we consider that there exist two campaigns on the platform, and there are two segments of customers. In other words, each arriving customer can have one of two heterogeneous valuations toward the two available campaigns in the platform. This case is of significant importance in modeling the revenue because it is inevitable not to take into account the fact that each person has a different taste and their valuations of even the same product could be totally different. We want to make it crystal clear that considering this important aspect of heterogeneous valuations will affect the revenue model, and as a result, the maximum platform's revenue itself. We will show how much the platform will lose in revenue if it does not take into account this single fact. In order to show different segments of customers in the model, we define two parameters, $\lambda_{1}$ and $\lambda_{2}$, which act for as the probability that an arriving customer belongs to what customer. For example, $\lambda_{1}$ is the probability that the arriving customer behaves like those people in segment 1. Furthermore, we consider 4 different parameters for $y$'s and 2 different parameters for $F$'s. The interpretation of each of these parameters are as follow:

$y_{ij}$: the intrinsic value, for which a customer in segment $j$ has of campaign $i$, $i = 1,2, j=1,2$\\
$F_{i}$: the remaining fund of campaign $i$ to reach its goal

Also, we include the network effects parameters in the model in this case. We will examine how the network effects could affect the purchase of customers in both segments. In this revenue model, we incorporate the network effect by using for parameters, i.e, $\alpha_{ij}$ for $i=1,2$ and $j=1,2$, which represent the impact of network effects of product $i$ on any arriving customers in segment $j$. 

Finally, we want to show how the classification problem can help us in the prediction of the best assortments when we consider the heterogeneous valuations, and whether the platform can trust the predictions of this model to provide an assortment of each arriving customer or not. In doing so, we will obtain the error rate and the percentage of loss in revenue for the classification model in this case.

Based on the above interpretation of the parameters in this case,the revenue model can be simplifies and written as below: 
\[
q_{ij} = \frac{\exp\left(y_{i} -F_{i}+\alpha_{i}\sum_{j=1}^{m}\lambda_{j}q_{ij}\right)}{1+\exp\left(y_{i} -F_{i}+\alpha_{i}\sum_{j=1}^{m}\lambda_{j}q_{ij}\right)},~ G\subseteq\lbrace 1,2\rbrace, i\in\lbrace 1,2\rbrace, m =2
\]
 and the optimal revenue is obtained from
 \[
 W^{*} = \max_{G}\left(\frac{a+b}{2}\right)(\xi+\omega)\sum_{j=1}^{m}\sum_{i\in G}  q_{ij}, m =2.
\]
Now we explain how he data has been driven by implementing the above model and using \textbf{Lamma}\ref{lemma1}. First, we generated 1000 rows of random numbers between 0 and M, where each row for $y$'s and $\alpha$'s has 2 columns (because we have 2 segments of customers), and each row in $F$'s has 1 column. For each nonnegative odd integer $k$, rows $j$ and $j+1$ correspond to product 1 and 2, respectively. Also, we generated 500 rows of random numbers for $\lambda$'s, and each one of the rows has two columns to represent two different segments of customers. We use \textbf{Lemma}\ref{lemma1} to compute the probability of purchase for each segment of customers and then calculating the revenues of all possible assortments which we can offer to each arriving customer. One important point that we made, in this case, is that the assortment that we offer to a customer in segment 1 could be different from the one we make available for customers in segment 2. The best assortments for segment 1 and segment 2 are the ones that bring the maximum revenue among the computed revenues. For having access to the data, please see "data4.m". 

Nest step is to implement the classification problem for this case. We separated the test and the training data into two different groups of data, i.e., one group for product 1 and the other for product 2. The multivariate regression has been implemented with the intercepted parameters and has been solved with CVX. There are 4 vectors of parameters that act for any pairs of $(i,j)$ of segments of customers and products. To put it simply, we define $\beta^{ij}$ be the parameters of the multivariate regression correspond to product $i\in\lbrace 1,2\rbrace$ and segment $j\in\lbrace 1,2\rbrace$. Hence, any response vector obtained from the multivariate regression is a vector of $4$ by $1$. To make these vectors to a vector of 0's and 1's and understand which product should be offered to what segments, we computed the norm $l_{2}$ of each response vector to all possible vector of $4$ by $1$ with 2 zeros and 2 ones, for example, $(1,1,0,0)$, which means product 1 should offer to customers both in segments 1 and 2 in this example.

The following are the values of these parameters in this case of the problem.

\[
\beta^{11} = \left[\begin{array}{cc}
-0.1645\\
0.0096\\
0.0031\\
0.6195\\
0.0033
\end{array}\right],~
\beta^{12} = \left[\begin{array}{cc}
0.8825\\
0.0002\\
0.0021\\
-0.6801\\
-0.0009
\end{array}\right],~
\beta^{21} = \left[\begin{array}{cc}
0.6738\\
0.0002\\
-0.0030\\
-0.6420\\
0.0017
\end{array}\right],~
\beta^{22} = \left[\begin{array}{cc}
-0.1184\\
0.0144\\
0.0016\\
0.5933\\
-0.0025
\end{array}\right]
\] 
The error rate and the average percentage of loss in revenue, in this case, are $0.8$ and $42.78\%$. These values and the results in Case 1 reveal the fact that the classification problem has a poor performance when the number of segments of customers increases, or put another way, the classification problem cannot give us a good predication of the best assortment when there exists heterogeneous behavior in the revenue model. That make sense, because this aspect of the problem increase the dimension of the classification problem.

For the revenue problem in this case we have $\max(r_{a})$, $\min(r_{a})$, and the average of $r_{a}$ equal $0.8537$, $2.4603e-30$, and $0.3580$, respectively. Comparing this part with Case 1. Part 1., we observe that the consideration of the heterogeneous behavior of customers towards different assortments is beneficial and brings more revenue for the platform, although it makes the problem more complicated and computationally expensive.

Although the numerical study includes these 4 cases, we want to make it clear to the reader that our models and codes can be generalized to any extends of number of segments and products, and we will end up with the results which we described in deep detail in this section 

\section{Conclusion}

In this paper, we studied crowdfunding platforms as some innovative business models which could be defined in terms of their revenue models. In this regard, we mentioned that considering the revenue maximization would be essential for these platforms in order to survive in the market. As discussed in the paper, these platforms charge various values among which the service fee is the only adjustable part by the platform owners. This fee could be defined as a function of the various parameters such as the assortment offered in each page of the website.
\\
\\
Based on the existing literature, an optimal assortment problem with random parameters has been proven to be in the class of Np-complete problems. In this case, we may assume that some small instances of the problem might be solved in polynomial time. For this purpose, we decided to think about a pseudo polynomial algorithm which might help to solve the problem in polynomial time. We believe that this result might be achievable if we extend our model and try to solve it by dynamic programming. 
\\

We began by discussing one version of the problem by considering a static deterministic model. We derived some optimal assortment policies under some conditions. In the numerical analysis section, we were able to observe some patterns while changing the values of different parameters of the model. We were also able to understand the importance of factors such as network effects and heterogeneous valuations of customers towards different campaigns and discuss how much revenue the platform could lose simply by not considering these factors. We also observed a dramatic change as the number of campaigns in every single assortment changes, and consequently, we expect the platform could gain more revenue when the number of campaigns in each assortment increases. Therefore, it is essential for the platform to use all its capacity on the website to make more campaigns available to each arriving customer. Two machine learning methods that we tried in this section were the multivariate regression and classification problem. We used our driven data to provide some predictions of the best assortments for the platform and realized that as the number of products and the segments of customers increases, these two problems have poor performance, and as a result, the platform cannot obtain its maximum revenue only by relying on the assortments provided by these two methods. However, As long as the number of products and segments of customers is low enough, the predication is acceptable and reliable, and the platform can obtain a reasonable amount of the maximum revenue by utilizing them.
\\
\\
We believe what has been discussed in this paper could be beneficial as a basis for some relative future research. First of all, we showed numerically that the previous results in the literature are not satisfactory and logically proper while dealing with a crowdfunding setting. Second, we proposed a multinomial logit model to predict customers' behavior by incorporating influential factors in this setting. Then, we solved some instances of the problem to provide some intuitions about the structure of the optimal policy. 
\\
\\
It is still an open question to us how we can find the optimal assortment while considering all of the existing uncertainties( i.e random parameters) and during a specific time horizon. Moreover, reading the relative literature we believe that modeling the same problem with a dynamic nature could be one future direction. While there exist a lot of uncertainties, as well as time horizon associated with each campaign, learning from the past experience (given an assortment), would be beneficial to estimate customers' behavior in the future. 
\\
\\
\section{Future Research}
An outline of the potential questions and future research topics could be asserted in the following way:
\begin{itemize}

    \item  Knowing that to creators, the value of a platform increases with the number of funders, and to funders, the value of a platform increases with the number of creators and other funders, how could we come up with some strategies in order to attract more backers and project owners?
    
    \item  Due to the fact that funding is highly skewed in crowdfunding platforms, how can we enhance these platforms?
    
    \item how to align incentives between platform, creators and funders?
    
    \item Economists study consumer behavior and how consumers continually make choices among products and services. They examine advantages of crowdfunding such as practicing menu pricing extracting a larger share of the consumer surplus, and disadvantages of crowdfunding such as constraining the choices of prices to attract a large number of funders. Management scholars find crowdfunding eliminates the effects of distance from funders whom creators did not previously know. Collecting more data and do a deeper analysis of motivations. In this analysis, we will examine factors that may influence motivations, such as domains and professional expertise. Planing to investigate how the same individual can participate in three distinct roles including observer, funder, and creator. While individuals initiate participation in crowdfunding in one role, initial evidence suggests that they transition between roles. Considering how participants learn innovation skills through crowdfunding platforms. Innovation leads to economic and social prosperity and we need people to have the skills, attitudes, and expertise necessary to innovate. 
    
    \item Considering the individual strategies people use to engage in crowdfunding. Initial evidence suggests that creators and funders extensively rely on social media to spread awareness of activity and promote engagement. Considering design principles to enhance creator and funder participation on crowdfunding platforms to ensure broader participation and encourage a variety of projects to get out into the world. We will possibly investigate including recommendation features, offering users selective choices and alternative mechanisms for financial and support, simplifying messaging, and identifying as a full service provider.
    
    \item Any arrival to the system plays one of the roles of an observer, a campaign owner and a backer. Taking into account this and the fact that the platform's revenue highly depends on the number of backers and campaign owners, how can the platform tract past and current observers to encourage them to become future backers or campaign owners? 
    \item One research opportunity in the area of operations research with crowdfunding settings is to solve the problems, which every campaign owner involves with. Some of the campaign owners' concerns are to make a good business model, production planning and delivery systems as well as to set profitable prices for their offered perks. In addition, successful campaigns which fail to deliver their products need policies to overcome their failures. 
    \item There are also some other research topic when we see the problem with the backer's prospective. One of the recent research topics which has been done in this area is the project's likelihood of success (M. Hu et al. 2017). We think another stimulating topic is to investigate whether and when project creators will successfully deliver the products.
\end{itemize}

\bibliographystyle{unsrt}

\end{document}